\documentclass[16pt,twocolumn,showkeys,showpacs,preprintnumbers,amsmath,amssymb]{revtex4}
\usepackage{graphicx}
\usepackage{dcolumn}
\usepackage{bm}
\usepackage{amsmath}
\usepackage{amssymb}
\usepackage{amsmath,amsthm,amssymb,amscd}
\usepackage{latexsym}
\usepackage{indentfirst}
\usepackage{subfigure}
\usepackage{graphicx}

\makeatother
\theoremstyle{plain}

\begin{document}

\date{\today}

\title{
  \bf Loose mechanochemical coupling of molecular motors
}

\author{Yunxin Zhang}\email[Email: ]{xyz@fudan.edu.cn}
\affiliation{Shanghai Key Laboratory for Contemporary Applied Mathematics,
Centre for Computational System Biology,\\
School of Mathematical Sciences, Fudan University, Shanghai 200433, China.
}

\begin{abstract}
In living cells, molecular motors convert chemical energy into mechanical work. Its thermodynamic energy efficiency, i.e. the ratio of output mechanical work to input chemical energy, is usually high. However, using two-state models, we found the motion of molecular motors is loosely coupled to the chemical cycle. Only part of the input energy can be converted into mechanical work. Others is dissipated into environment during substeps without contributions to the macro scale unidirectional movement.
\end{abstract}



\maketitle

{\bf Introduction.}
In biological cells, molecular motors are individual protein
molecules that are responsible for many of the biophysical
functions of the cellular movement and mechanics \cite{Howard2001}. They work in
nanometer range and convert chemical energy, stored in ATP molecules,
into mechanical work. Important examples of molecular motors are kinesin, dynein
\cite{Gennerich2009} and mysion \cite{Sakamoto2008, Vale2003}.
Recent experimental data indicate that the energy efficiency of molecular motors is high. 

In literatures, there are many theoretical models to study molecular motors, such as Fokker-Planck equation \cite{Zhang20091}, Langevin equation \cite{Reimann2009}, lattice model \cite{Kolomeisky2007}, network model \cite{Liepelt2007} and etc. All of the existing models can be roughly classified into two categories: (1) There is only one chemical state, or equivalently only one tilted periodic potential in the model. (2) There are multiple chemical states, i.e. there are several periodic potentials in the model. To some extents, any one model of the second class is equivalent to one of the first class \cite{Wang2006}. Usually the first class models are simple, and can be used to get explicit formulations of important biophysical quantities, such as the mean velocity \cite{Derrida1983, Fisher1999}, effective diffusion coefficient \cite{zhang20092} and mean first passage time \cite{Pury2003}. However, due to the oversimplification, more reasonable biophysical properties of molecular motors can not be obtained from them. On the other hand, though the second class models seem more reasonable, it is too difficult to obtain explicit results \cite{Wang2004, Lipowsky2000}. The simplest model of class two is the two-state model, it has been frequently employed by many authors \cite{Chen1999, Bier1993, Frank1995, Prost1994}. But most of their results are based on numerical calculations and then difficult to do further analysis. In this research, we will give some explicit results of the two-state model, including the mean velocity and energy efficiency. The main conclusion that we draw from the two-state model is that, the motion of molecular motors is loosely coupled to the chemical cycle, its mean velocity might be zero even if there exists nonzero input energy, which is in accordance with the recent studies \cite{Yildiz2008, Masuda2009, Gao2006, Gerritsma2009}.

{\bf One-state models.}
The motor motion can be modeled by the Langevin equation
\begin{equation}\label{eq1}
\xi \dot{x}(t)=-\partial_x \Phi(x)+\sqrt{2k_BT\xi}f(t),
\end{equation}
where $f(t)$ is Gaussian white noise, $\xi$ is viscous friction coefficient, $k_B$ is Boltzmann's constant, $T$ is the absolute temperature, and $\Phi(x)$ is a tilted periodic potential, i.e. $\Phi(x)-\Phi(x+L)\equiv\Delta\Phi$ is constant (the step size $L=8$ for conventional kinesin and cytoplasmic dynein, and $L=36$ for Myosin V). Or equivalently, it can be modeled by the following Fokker-Planck equation
\begin{equation}\label{eq2}
\partial_t \rho=\partial_x\left(\partial_x\Phi\rho/\xi+D\partial_x \rho\right),
\end{equation}
where $D=k_BT/\xi$ is free diffusion coefficient, $\rho(x,t)$ is the probability density of finding the motor at time $t$ and position $x$. Using either (\ref{eq1}) or (\ref{eq2}), the steady state mean velocity can be obtained as follows
\begin{equation}\label{eq3}
V=\frac{\left(1-\exp(-\beta\Delta\Phi)\right)DL}{\int_0^L\exp(-\beta\Phi(x))
\left(\int_x^{x+L}\exp(\beta\Phi(y))dy\right)dx},
\end{equation}
with $\beta=1/k_BT$. Obviously $V>0$ if and only if $\Delta\Phi>0$. If there exists external load $F_{ext}$, the potential $\Phi(x)$ should be replaced with $\phi(x)=\Phi(x)-F_{ext}x$.

Another popular one-state model is the one-dimensional hopping model: in which the motor in mechanochemical state
$j$ can jump forward to state $j+1$ with rate $u_j$, or backward to state $j-1$ with rate $w_{j}$. After
moving $N$ states forward or backward the motor comes to the same chemical state but spatially shifted by a step size $L$. Using this one-dimensional hopping model, the mean velocity of the molecular motors is
\begin{equation}\label{eq4}
V_N=\frac{L\left[1-\prod_{j=0}^{N-1}\frac{w_i}{u_i}\right]}{\sum_{j=0}^{N-1}\left(\frac{1}{u_j}\left[1+\sum_{k=1}^{N-1}\prod_{i=j+1}^{j+k}\frac{w_i}{u_i}\right]\right)}. \end{equation}
Meanwhile, the free energy difference in one mechanochemical period is
$\Delta\mu=k_BT\ln\left(\prod_{j=0}^{N-1}{w_i}/{u_i}\right)$.
Obviously $V_N>0$ if and only if $\Delta\mu>0$. Actually, it can be proved mathematically that $\lim_{N\to \infty}V_N=V$ and $\Delta\mu=\Delta\Phi$ ($\Delta\mu=\Delta\phi$ if there exists external force, see \cite{Zhang2010}). By (\ref{eq3}) or (\ref{eq4}), the stall force of molecular motor can be easily obtained $F_s=\Delta\Phi/L$, and the thermodynamic energy efficiency is $\eta=F_{ext}L/\Delta\Phi$, which increases with the external load $F_{ext}$, and $\eta({F_s})=1$. Therefore, the one-state models are tightly mechanochemical coupled models. The maximum of energy efficiency is 1, which is attained at stall force $F_s$.

\begin{figure}
  \includegraphics[width=220pt]{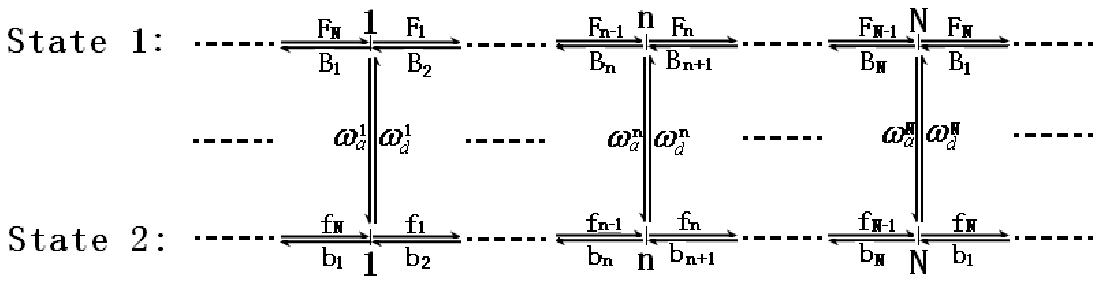}\\
  \includegraphics[width=220pt]{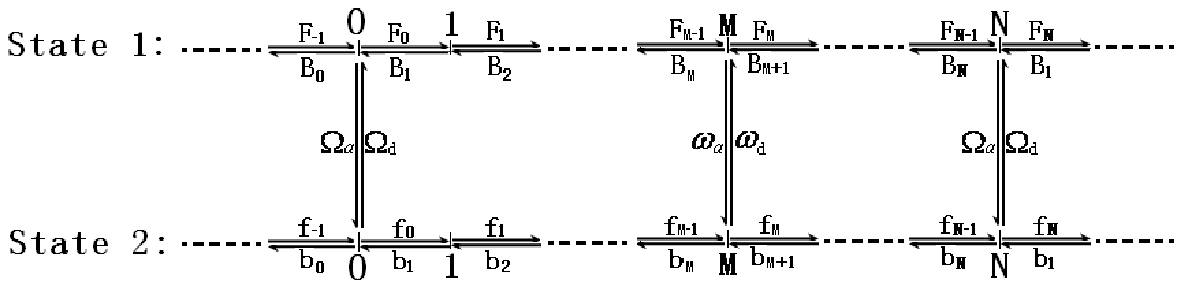}\\
  \caption{Schematic depiction of the two-state lattice models: ({\bf Up}) general case, ({\bf Down}) special case, in which  $\omega^i_a=\omega^i_d=0$ for $i\ne M, N$.}\label{Fig1}
\end{figure}

{\bf Two-state models.}
The general two-state continuous model is the following mechanochemical coupled Fokker-Planck equations:
\begin{equation}\label{eq6}
\left\{\begin{aligned}
\partial_t P=&D\partial_x(\beta
\Phi_1' P+\partial_x P)+\omega_d(x) \rho-\omega_a(x) P\cr
\partial_t \rho=&D\partial_x(\beta
\Phi_2' \rho+\partial_x \rho)-\omega_d(x) \rho+\omega_a(x) P,
\end{aligned}\right.
\end{equation}
with $0\le x\le L$. Where $P(x,t)$ and $\rho(x,t)$ are the probability densities of finding motor at time $t$, position $x$ and in states 1 and 2 respectively. $\omega_d(x), \omega_a(x)$ are transition rates between the two chemical states at position $x$.
The corresponding general two-state lattice model is as follows (see Fig. \ref{Fig1})
\begin{equation}\label{eq7}
\begin{aligned}
&\left\{\begin{aligned}
\frac{d}{dt}P_n=&F_{n-1}P_{n-1}-(F_n+B_n)P_n\cr
&+B_{n+1}P_{n+1}-\omega_n^aP_n+\omega_n^d\rho_n\cr
\frac{d}{dt}\rho_n=&f_{n-1}\rho_{n-1}-(f_n+b_n)\rho_n\cr
&+b_{n+1}\rho_{n+1}+\omega_n^aP_n-\omega_n^d\rho_n,
\end{aligned}\right.\end{aligned}
\end{equation}
with $n=1,2,\cdots, N$, $P_n$ and $\rho_n$ are probabilities in states 1 and 2 respectively. All the rates, $F_n, B_n, f_n, b_n, \omega_n^a, \omega_n^d$, are periodic with period $N$. In the following, we will only discuss the two-state lattice model (\ref{eq7}), but all the corresponding results also can be obtained by the two-state continuous model (\ref{eq6}) \cite{Zhang20098}.

Although the steady state solutions of the general two-state lattice model
(\ref{eq7}) can be obtained explicitly, it will be certainly easier for us to show
our conclusions only by employing one special case, in which
$\omega^i_a=\omega^i_d=0$ for $i\ne M, N$.
For the simplicity of notations, we denote $\omega^N_a, \omega^N_d$ by $\Omega_a,
\Omega_d$ and  $\omega^M_a, \omega^M_d$ by $\omega_a, \omega_d$ respectively (see
Fig. \ref{Fig1}). To this special case, one can calculate that the probability flux is
\begin{equation}\label{eq9}
\begin{aligned}
J=&\left[J_c\left(\frac{S_N}{T_N}-\frac{s_N}{t_N}\right)+\frac{(R_N-1)W}{T_N}+\frac{(r_N-1)U}{t_N}\right]\frac{P_N}{W},
\end{aligned}
\end{equation}
in which $S_k=\left.\left[\sum_{i=M+1}^{k}\prod_{j=i}^{k}({F_j}/{B_j})\right]\right/F_k$,
$T_k=$ $\left.\left[\sum_{i=1}^{k}\prod_{j=i}^{k}({F_j}/{B_j})\right]\right/F_k$,
$R_k=\prod_{i=1}^k({F_{i-1}}/{B_i})$,
$J_c=(\omega_a\Omega_dR_M-\Omega_a\omega_dr_M)$,
$U=\Omega_a+\omega_aG_M-\omega_dh_M$, $W=\Omega_d+\omega_dg_M-\omega_aH_M$
and $P_N=W\left/\left(\sum_{k=1}^N\left[(G_k+h_k)W+(g_k+H_k)U\right]\right)\right.$
is the probability of finding motors in state $1$ and at position $N$.
$$
\begin{aligned}
G_k=&\left\{\begin{array}{ll}
R_k-\frac{T_k}{T_N}(R_N-\Omega_aS_N-1)  &1\le k\le M\cr
R_k-\Omega_aS_k-\frac{T_k}{T_N}(R_N-\Omega_aS_N-1) &\textrm{otherwise}
\end{array}\right.\cr
H_k=&\left\{\begin{array}{ll}
-\frac{T_k}{T_N}\Omega_dS_N\qquad &1\le k\le M\cr
-\frac{T_k}{T_N}\Omega_dS_N+\Omega_dS_k & \textrm{otherwise}
\end{array}\right..
\end{aligned}
$$
Similar expressions for $r_k, s_k, t_k, g_k, h_k$ can be obtained by replacing $F_i, B_i, \Omega_a, \Omega_d$ in expressions of $R_k, S_k, T_k, G_k, H_k$ with $f_i, b_i, \Omega_d, \Omega_a$ respectively.
Under no external load, the input energy in unit time is
\begin{equation}\label{eq10}
\begin{aligned}
\Delta\mu
=&k_BT\left[\frac{R_N-1}{T_N}V\ln R_N+\frac{r_N-1}{t_N}U\ln r_N\right]+\frac{k_BTJ_cP_N}{V}\cr
\times&\left(\frac{t_N-s_N}{t_N}\ln r_N-\frac{T_N-S_N}{T_N}\ln R_N+\ln\frac{\omega_a\Omega_dR_M}{\Omega_a\omega_dr_M}\right).\cr
\end{aligned}
\end{equation}
One can show that $\Delta\mu=0$ if and only if $R_N=r_N=1$ and $J_c=0$. $R_N=r_N=1$ means the potentials in states 1 and 2 are all periodic and so no input energy in either of the two states. $J_c=0$ means there is no input energy during transitions between the two states.

In the following, we assume that, under no external load the potentials in each states are all periodic since, biophysically the input energy to molecular motors comes from ATP molecules, which is hydrolyzed during transitions between different chemical states. So, under no external load, the mean velocity of the motor is
\begin{equation}\label{eq11}
V=JL=\left({S_N}/{T_N}-{s_N}/{t_N}\right)J_cL{P_N}/{W},
\end{equation}
and the input energy in unit time is
\begin{equation}\label{eq12}
\Delta\mu=k_BTJ_c{P_N}\left[\ln({\omega_a\Omega_dR_M}/{\Omega_a\omega_dr_M})\right]/{W}.
\end{equation}
Obviously, $\Delta\mu=0$ implies $V=0$, but $V=0$ does not give $\Delta\mu=0$. The reason is that there might exist nonzero circumfluence of probability (see Fig. \ref{Fig4}), i.e. the motors might make a forward substep in one state but stepping back in another one. In these processes, free energy is consumed but no effective steps are made.

under nonzero external load $F_{ext}$, the transition rates satisfy ${F_i(F_{ext})}/{B_{i+1}(F_{ext})}=[{F_i(0)}/{B_{i+1}(0)}]\exp({-{\delta^1_i\beta F_{ext}L}})$, ${f_i(F_{ext})}/{b_{i+1}(F_{ext})}=[{f_i(0)}/{b_{i+1}(0)}]\exp({-{\delta^2_i\beta F_{ext}L}})$, in which $\sum_{i=0}^{N-1}\delta^j_i=1$ ($j=1, 2$). Depending on the parameters  $\Omega_a, \omega_d, \omega_a, \Omega_d$, the motor might move to the right direction ($V>0$) or to the left direction ($V<0$). In our discussion, we always assume that the direction of the external load is opposite to the motor motion direction, i.e. the external load points to the left ($F_{ext}>0$) if $V>0$ and points to the right ($F_{ext}<0$) if $V<0$. This assumption means $F_{ext}V\ge 0$.

To the nonzero external load cases, the free energy difference in one
mechanical cycle is \cite{Zhang20098}
\begin{equation}\label{eq13}
\begin{aligned}
\Delta G
=&k_BT\left\{[(R_N-1)W/{T_N}-J_c(T_N-S_N)/{T_N}]\ln R_N\right.\cr
+&[(r_N-1)U/{t_N}+J_c(t_N-s_N)/{t_N}]\ln r_N\cr
+&\left.J_c\ln[{(\omega_a\Omega_dR_M)}/{(\Omega_a\omega_dr_M)}]\right\}{P_N}/{V}.
\end{aligned}
\end{equation}
It can be verified that the input energy $\Delta\mu(F_{ext})=\Delta G(F_{ext})+F_{ext}V$. In which $\Delta G(F_{ext})\ge 0$, and $\Delta G(F_{ext})=0$ if and only if $J_c=0$. Therefore, $\Delta G(F_{ext})\ge 0$ if and only if $\Delta\mu(F_{ext})=0$, which means $F_{ext}V\le \Delta\mu(F_{ext})$ and the equality holds if and only if $F_{ext}V=\Delta\mu(F_{ext})=0$.

The velocity-force relation can be obtained by formulation (\ref{eq9}) with the stall force $F_s$ is attained when $J=0$. It should be pointed out that, under stall force, $V=0$ but the input energy $\Delta\mu(F_{ext})\ne 0$ if $J_c\ne 0$, which only depends on the parameters $\Omega_a, \omega_d, \omega_a, \Omega_d$. In fact, $J_c=[\omega_a\Omega_dR_M(0)-\Omega_a\omega_dr_M(0)]\exp(-\delta\beta F_{ext}L)$ with $\delta=\delta_0^j+\cdots+\delta_{M-1}^j$ ($j=1, 2$). Therefore, the motion of molecular motors is loosely coupled to the chemical cycle, this is consistent with the recent experimental results \cite{Yildiz2008, Masuda2009, Gao2006, Gerritsma2009}.

The flashing rachet model can be regarded as one of the special cases of the above models: in which one of the potentials (for example, in state 2) is constant, i.e. $f_i(0)=b_i(0)\equiv f$ ($i=1, 2, \cdots, N$). To this much special case, corresponding results can be derived easily.

{\bf Limit properties and energy efficiency.}
To some molecular motors, it might be possible to change the transition rates $\Omega_a, \omega_d, \omega_a, \Omega_d$, which usually depend on temperature, ATP concentration and potential profiles. For simplicity, we suppose $(\Omega_a, \omega_d, \omega_a, \Omega_d)$=$\lambda(\Omega^0_a, \omega^0_d, \omega^0_a, \Omega^0_d)$ and $\lambda$ is a parameter that can be changed experimentally.

One can easily show that the mean velocity $V\to 0$ with $\lambda\to 0$. But for $\lambda\to \infty$, the mean velocity tends to
$$\left.\left[{\Psi_1\left(\frac{S_N}{T_N}-\frac{s_N}{t_N}\right)+\Psi_2\left(\frac{R_N-1}{T_N}+\frac{r_N-1}{t_N}\kappa_2\right)}\right]L\right/{\Xi}$$ where $\Psi_1=\kappa_1 R_M-\kappa_2 r_M$, $\Psi_2=\kappa_1{T_MS_N}/{T_N}+{t_Ms_N}/{t_N}$, $\kappa_1={\omega^0_a}/{\omega^0_d}$, $\kappa_2={\Omega^0_a}/{\Omega^0_d}$ and
$\Xi=\Psi_2\sum_{k=1}^N\left(R_k+\kappa_2r_k\right)$+$\Psi_1\left[\sum_{k=1}^N\left({t_ks_N}/{t_N}-{T_kS_N}/{T_N}\right)\right.$
+$\left.\sum_{k=M+1}^N(S_k-s_k)\right]$. Similarly, as $\lambda\to \infty$, the input
energy $\Delta\mu$ tends to
$k_BT\Psi_1\ln({\kappa_1R_M}/{\kappa_2r_M})/\Xi$. So, the output power of
molecular motor has a limit which only depends on the parameters
$\kappa_1, \kappa_2$ (i.e., is independent of $\lambda$).

\begin{figure}
  \includegraphics[width=120pt]{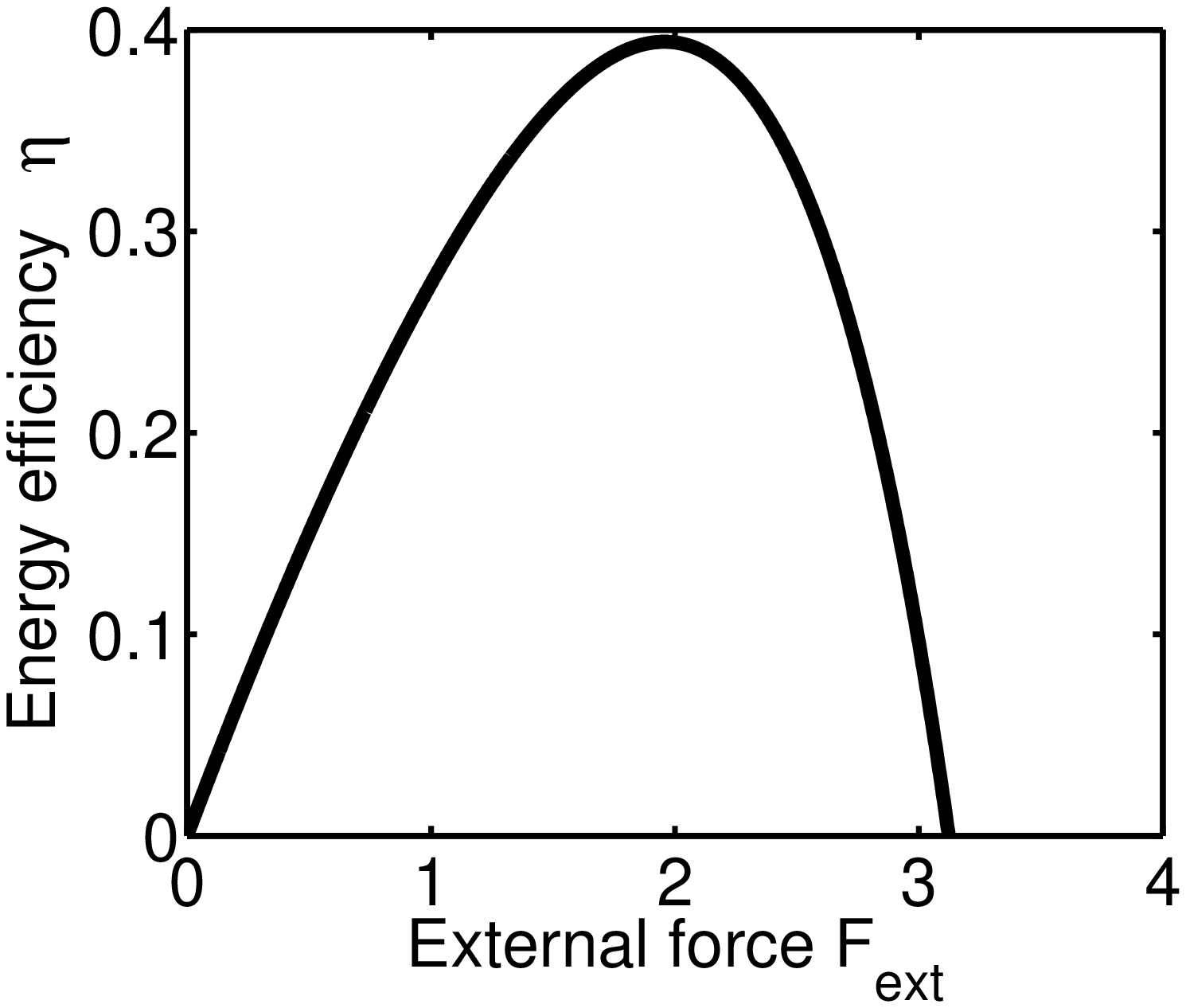}\includegraphics[width=120pt]{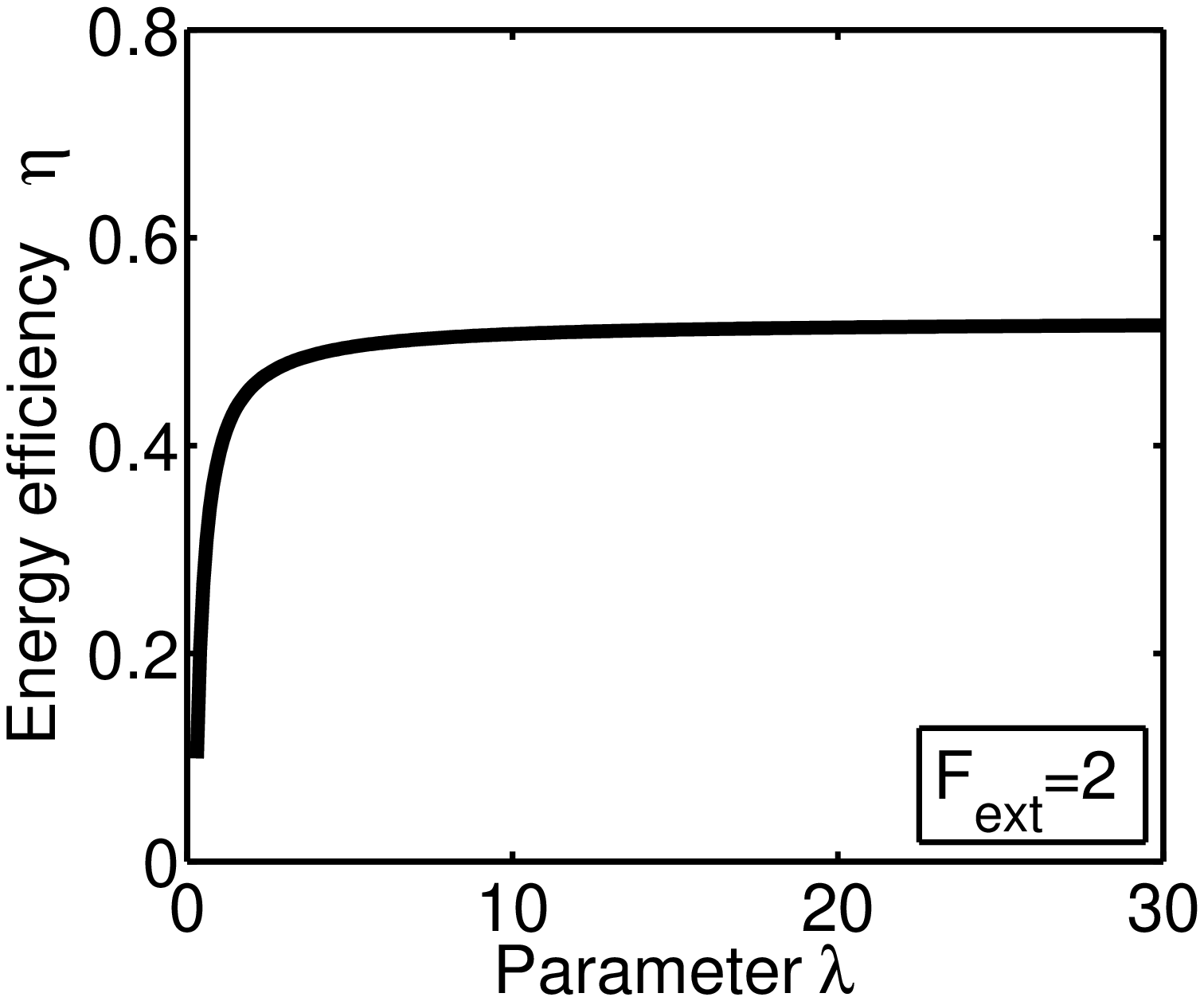}\\
  \caption{The thermodynamic energy efficiency of molecular motor calculated by
  the special two-state model, as a function of external load $F_{ext}$
  ({\bf left}) and as a function of parameter $\lambda$ ({\bf right}), which is defined by
  $(\Omega_a, \omega_d, \omega_a, \Omega_d)$=$\lambda(\Omega^0_a, \omega^0_d, \omega^0_a, \Omega^0_d)$.
  The parameters used in the model are $N=2, M=1, k_BT=1, F_0(0)=100, B_1(0)=0.5, F_1(0)=0.005, B_0(0)=1, f_0(0)=0.5, b_1(0)=1, f_1(0)=50, b_0(0)=25, L=1, \Omega_a=20, \Omega_d=5, \omega_a=10, \omega_d=50, \delta^j_1=0.9,  \delta^j_2=0.1$ (j=1, 2). 
  }\label{Fig6}
\end{figure}

\begin{figure}
\includegraphics[width=225pt]{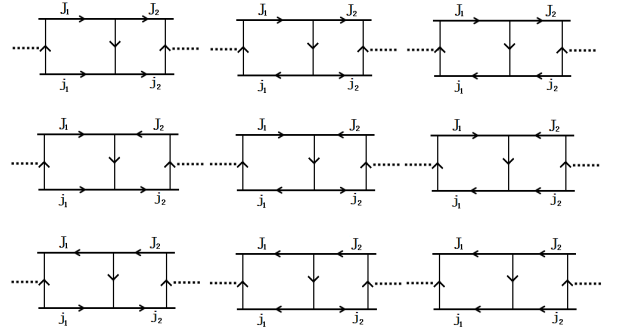}\\
  \caption{Different cases of probability flux in which there are only two nonzero transition rates.}\label{Fig4}
\end{figure}

Another interesting biophysical property of molecular motor is its energy
efficiency. By (\ref{eq9}) (\ref{eq12}), its thermodynamic energy efficiency
$\eta(F_{ext}):={FV}/{\Delta\mu}$ is
\begin{equation}\label{eq14}
\begin{aligned}
\eta=\frac{\left[J_c\left(\frac{ S_N}{ T_N}-\frac{ s_N}{ t_N}\right)+\left(\frac{R_N-1}{T_N}V+\frac{r_N-1}{t_N}U\right)\right]F_{ext}L}{J_ck_BT[\ln({\kappa_1  R_M}/{\kappa_2  r_M})]}
\end{aligned}
\end{equation}
Obviously, $\eta(0)=\eta(F_s)=0$. From the loose mechanochemical coupling discussion in the above section, we can easily know $\eta<1$.
With $F_{ext}$ increases from zero to $F_s$, the mean velocity decreases to zero monotonically. But for the efficiency $\eta$, there exists a maximum value between 0 and $F_s$ (see Fig. \ref{Fig6}).

{\bf Discussions.}
The analysis of two-state models indicates the motor motion is loosely coupled to the chemical cycle. The biophysical reason of this loose mechanochemical coupling is there exist forward and backward substeps, which had been found experimentally in \cite{Block2003, Coppin1996}. Free energy is consumed in each substeps, but only part of them really contributes to the effective unidirectional motion. In the two-state models, this corresponds to the existence of probability circumfluence. Due to the temporal symmetry of states 1 and 2, there are altogether 9 different cases of the probability flux (see Fig. \ref{Fig4}). Free energy is consumed but without effective forward motion in each circumfluence. To the general two-state lattice models, there exist much more cases of the probability circumfluence.

\begin{figure}
  \includegraphics[width=110pt]{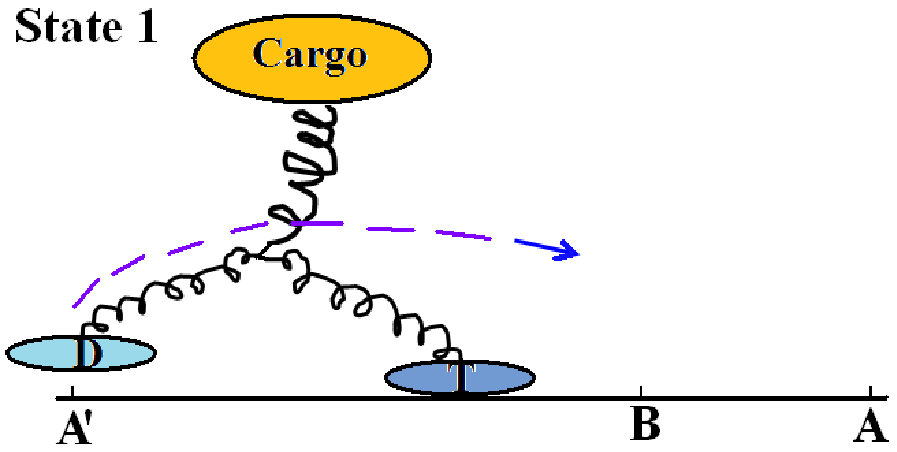}\includegraphics[width=110pt]{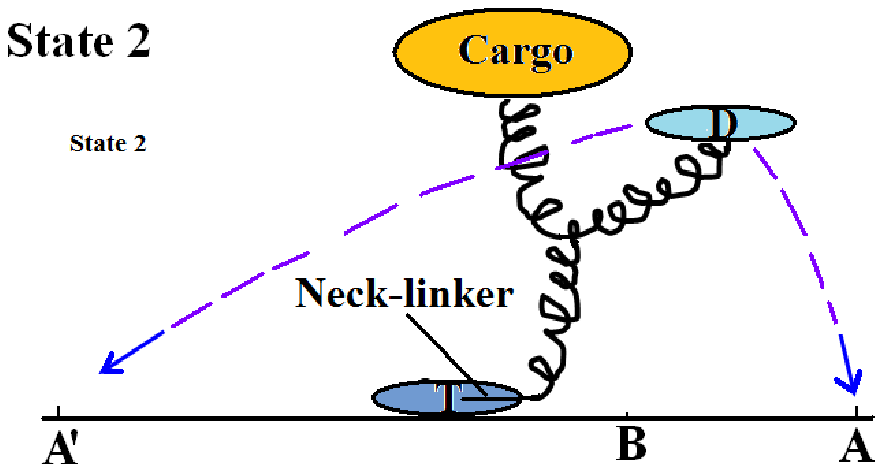}\\
  \includegraphics[width=130pt]{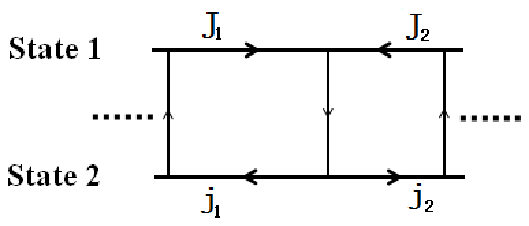}\includegraphics[width=100pt]{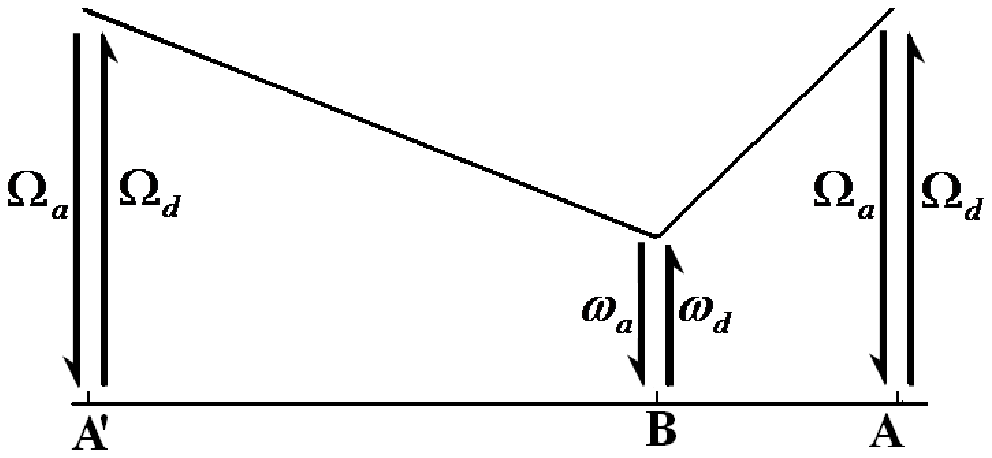}\\
  \caption{Two-state model of motor protein kinesin: In state 1, the motor moves from location $A$ or $A'$ to the potential well $B$, in state 2, the motor moves stochastically from location $B$ to $A$ or $A'$. The mechanical stepsize $|A'A|$ is about 8 nm, and the substep $|A'B|$ is about 5 nm \cite{Zhang2008}.}\label{Fig5}
\end{figure}

Using special cases of the two-state continuous models (\ref{eq6}), the same conclusion can be obtained. For example, to the special case: $\omega_a(x)=\omega_d(x)\equiv 0$ for $0<x<a, a<x<L$, the explicit expression of the velocity gives: the mean velocity $V>0$ if and only if $\left[\Omega_a\omega_de^{\beta(\Phi_2(0)-\Phi_2(a))}-\omega_a\Omega_de^{\beta(V_1(0)-\Phi_1(a))}\right]\times$ $\left[{\int_a^Le^{\beta \Phi_1(y)}}/{\int_0^ae^{\beta \Phi_1(y)}}-{\int_a^Le^{\beta \Phi_2(y)}}/{\int_0^ae^{\beta \Phi_2(y)}}\right]<0$. But the input energy is always positive. Therefore, the fact of loose mechanochemical coupling also can be found from this continuous models. As an example, the motion of motor protein kinesin can be schematically described by the special case of our two-state model (see Fig. \ref{Fig5}). In which the transition from state 1 to state 2 is due to the neck linker docking of the microtubule bounded head (leading head), the transition from state 2 to state 1 is due to ATP hydrolysis and phosphate release. Due to the existence of substeps, which might have no any contributions to the macroscopic unidirectional motion of kinesin, the energy efficiency of kinesin is certainly smaller than 1.


\vskip2cm

\acknowledgments{This study is funded by the Natural
Science Foundation of Shanghai (under Grant No. 11ZR1403700).}

\end{document}